\documentclass[twocolumn,superscriptaddress,amsmath,amssymb,aps]{revtex4-1}
\usepackage{physics}
\usepackage{graphicx}
\newcommand{\pin}[0]{\ket{\ket{\textbf{p}_{\nu_l}(t_1)}}}
\newcommand{\pfin}[0]{\ket{\ket{\textbf{p}_{\nu_l}(t_2)}}}
\usepackage{hyperref}
\hypersetup{
	colorlinks = true,
	urlcolor   = blue,
	linkcolor  = blue,
	citecolor  = blue
}
\begin{document}
\title{Multiclass classification of dephasing channels}
\author{Adriano~M. Palmieri} \email{adriano.palmieri@skoltech.ru}
\affiliation{Skolkovo Institute of Science and Technology, 121205 Moscow, Russia}
\author{Federico Bianchi}\email{federico.bianchi@unimib.it}
\affiliation{Bocconi University, I-20136 Milan, Italy.}
\author{Matteo G. A. Paris}\email{matteo.paris@fisica.unimi.it}
\affiliation{Quantum Technology Lab, Dipartimento di Fisica {\em Aldo Pontremoli}, Universit\`a degli Studi di Milano, I-20133 Milano, Italy}
\affiliation{Istituto Nazionale di Fisica Nucleare - Sezione di Milano, I-20133 Milano, Italy}
\author{Claudia Benedetti} \email{claudia.benedetti@unimi.it}
\affiliation{Quantum Technology Lab, Dipartimento di Fisica {\em Aldo Pontremoli}, Universit\`a degli Studi di Milano, I-20133 Milano, Italy}
\begin{abstract}
We address the use of neural networks (NNs) in classifying the 
environmental parameters of single-qubit dephasing channels. In particular, we investigate the performance of linear perceptrons and of two non-linear NN architectures.
At variance with time-series-based approaches, our  goal is to  learn a discretized probability distribution over the  parameters  using  tomographic data at just \emph{two} 
random instants of time. We consider dephasing channels originating either from classical $1/f^{\alpha}$ noise or from the interaction with a bath of quantum oscillators.  The parameters to be classified are the {\em color} $\alpha$ of the classical noise or the {\em Ohmicity} parameter $s$ of the quantum  environment. In both cases, we found that NNs are able to exactly classify parameters into  16 classes using noiseless data (a linear NN is enough for the color, whereas a single-layer NN is needed for the Ohmicity). In the presence of noisy data (e.g. coming from noisy tomographic measurements), the network is  able to classify the color of the $1/f^{\alpha}$ noise into 16 classes with about $70\%$ accuracy, whereas classification of Ohmicity  turns out to be challenging. We also consider a more {\em coarse-grained} task, and train the network to discriminate between two macro-classes corresponding to $\alpha \lessgtr 1$ and 
$s \lessgtr 1$, obtaining  up to $96 \%$ and   $79 \%$ accuracy using single-layer NNs.
\end{abstract} \date{\today}
\maketitle
\section{Introduction}
Noisy intermediate scale quantum devices are currently available \cite{Zhong1460}, but the technology does need to overcome this intermediate stage to allow their full scalability \cite{acn2018}. In this endeavour, the ability to  fully characterize the unavoidable impact 
of the  environment on the quantum system of interest is a cornerstone for the implementation 
of quantum information tasks \cite{breuer,qcon}, including redundancy calibration or high 
fidelity quantum gates. In particular, high performance quantum simulators \cite{Johnson2014}, sensors \cite{rossi2020noisy}, and computers \cite{Garcia-Perez2019} 
rely on the precise knowledge of the surrounding environment \cite{Foti2019}, which is usually modelled as a complex system, made of many and non-controllable degrees of freedom. 

Dephasing channels describe the open dynamics of quantum systems that do not exchange 
energy with their surrounding environments, but are nevertheless affected in their coherence \cite{apl17}. If a qubit goes through a dephasing channel, the environment, and in 
particular the noise spectral features, strongly affects the qubit reduced dynamics. 
Upon accessing the quantum state of the qubit it is thus possible to infer the spectral 
parameters of the noise. The use of a single qubit as a quantum probe for environmental 
parameters has been extensively analyzed \cite{benedetti14, benedetti2014, benedetti18, PhysRevX.10.011018,Tamascelli20, gianani20}. 
However, in all these studies the qubit probe must be measured at a very 
specific time in order to achieve high precision. In this paper, we go beyond this 
constraint, and design a classification technique which requires tomographic data 
at just two random instants of times, within a fixed time-window. 
 
We consider two paradigmatic models of dephasing. The first is the celebrated $1/f^{\alpha}$ classical noise originating from the interaction with a set of classical fluctuators \cite{paladino14}. The second, referred to as quantum dephasing noise, comes from the 
interaction of the qubit with a bath of bosonic oscillators  \cite{breuer}. 
The $1/f^{\alpha}$ noise is an ubiquitous model of noise, which affects both quantum and classical systems. The value of the  exponent $\alpha$ is usually referred to as the {\em color} 
of the noise, and depends on the details of the system under investigation \cite{kumar16, handel17, ag18, Bergli09, kazakov20}. If the environment is  a collection of bosonic modes, the spin-boson model is instead suitable to describe the dephasing dynamics of a single qubit interacting 
with the quantum bath. 

Machine learning (ML) for physics aims at discovering and implementing data driven-adaptive approaches to solve physical problems. ML is 
finding a widespread, growing number of applications in quantum physics including quantum tomography \cite{carleotorlai,carasquillarnn,processtomography,adrifede,GANoptics,quantumAE}, metrology and sensing \cite{Adaptivebayes,cimini19,cimini21} , entanglement classification \cite{bellsforclassification, paternostro1}, quantum thermodynamics \cite{reinforcementhermo}, quantum control\cite{RLquantumcontrol}, communication \cite{RLquantumcommunication}, and renormalization group. 
In turn, 
attention has been devoted also to open quantum systems dynamics and noise mitigation. 
Restricted Boltzmann machines and neural quantum states found several applications to 
dynamical systems \cite{carleo1,BANDYOPADHYAY,steadystate} undergoing dephasing, and more recently  autoregressive models over experimental data has been discussed \cite{RNN}. Other techniques 
have been also considered for simulation of open quantum systems, as convex-optimization for Lindbladian  \cite{fitting}, unsupervised tensor network learning \cite{processtomography}, adaptive regression strategy \cite{bilinear}, and deep evolutionary approaches \cite{evolutionarydnn}.
On the flip side, long short-term memory architecture has shown itself  able to offer high improvements to the arduous problems of noise mitigation and spectrum understanding \cite{LSTMnoisemitigation}.

In this paper, we address discrimination (i.e. classification) problems for 
single-qubit dephasing channels, originating either from the interaction with a 
classical or a quantum environment. Specifically, we want the network to distinguish 
among a discretized set of spectral parameters, i.e. the color of the noise for 
classical dephasing and the Ohmicity for the quantum baths.
Our approach relies on the fact that deep neural networks are effective at learning functions, or at replicating the dynamical parameters thereof
\cite{NNschroedingerequation,reservoirchaotic,rnnattractor,rnnmissingdata,reviewrnn}.
Our goal is to infer the environment parameters by feeding the network with  
tomographic data taken at (only) two random instants of time. In this way, we 
lift the constraint of the qubit state being measured at a specific time.
We show that NNs are able to perfectly discriminate between different values of the spectral parameters, as long as we use noiseless data.  Noisy data, instead, reduce the performance.  
More specifically, classification of the color of classical noise is still feasible, whereas 
boson baths are more difficult to classify. Higher levels of accuracy are still accessible 
by reframing the problem into a two-class discrimination task, which is of interest in several potential applications \cite{benedetti14,reinforcementhermo,HESABI2019377}. 

The paper is structured as follows: In section II we introduce the physical model for a dephasing qubit stemming either from a classical or a quantum environment; In section III we describe the dataset preparation and we briefly review the learning models salient characteristics and metrics. In Section IV we present our results for two main families of data: noiseless  and noisy data. We address accuracy and macro-F1 score to evaluate the classification performance. Section V closes the paper with  final discussions and remarks.

\section{The model}\label{sec:model}
A qubit dephasing channel is described by the quantum map:
\begin{equation}
\rho(t)=\frac{1+\Lambda(t)}{2}\rho(0)+\frac{1-\Lambda(t)}{2}\sigma_z \rho(0) \sigma_z
\label{channel}
\end{equation}
where $\sigma_z$ is a Pauli matrix, $\rho(0)$ is the initial density matrix of the qubit, 
and $\Lambda(t)$ is referred to as the dephasing function. The explicit expression 
of $\Lambda(t)$ depends on the specific microscopic model describing 
the system-environment interaction.  The dephasing map in Eq. (\ref{channel}) may originate
from different mechanisms. Here we
consider two of them, which are relevant to different branches of quantum technology, corresponding to situations where the qubit interacts with a classical fluctuating environment or with a bosonic bath \cite{rtn1,razavian19}.
If the qubit interacts with a  classical bistable fluctuator 
having random switching rates $\gamma$, 
the dephasing coefficient takes the expression \cite{benedetti114}:
\begin{align}
\Lambda_c(t,\alpha)=\int_{\gamma_1}^{\gamma_2} G(t,\gamma) p_{\alpha}(\gamma) d\gamma
\end{align}
with
\begin{align}
G(t,\gamma)=e^{-\gamma t}\left[ \cosh(\delta t)+\frac{\gamma }{\delta} \sinh(\delta t)\right],
\end{align}
$\delta=\sqrt{\gamma^2-4}$, and $t$ is a dimensionless time. 
If the probability distribution $p_{\alpha}(\gamma)$ is defined as:
\begin{align}
p_{\alpha}(\gamma)=\begin{cases}
\frac{1}{\gamma \ln(\gamma_2/\gamma_1)}&\alpha=1\\
\frac{\alpha-1}{\gamma^{\alpha}}\left[ \frac{(\gamma_1 \gamma_2)^{\alpha-1}}{\gamma_2^{\alpha-1}-\gamma_1^{\alpha-1}}\right]&\alpha\neq 1
\end{cases},
\end{align}
the overall spectrum of the noise corresponds to a $1/f^{\alpha}$ distribution. 
As mentioned above, the parameter $\alpha$ is referred to as the {\em color} of the noise, since it determines the weight of the different frequencies $f$.

If the qubit is interacting with a bath of quantum oscillators at zero temperature, the dephasing function has an exponential decaying form \cite{breuer}:
\begin{align}
\Lambda_q(t,s)=e^{-\Gamma(t, \omega_c,s)}
\end{align}
with
\begin{align}
\Gamma(t, \omega_c,s)=\begin{cases}
\frac12 \ln(1+\omega_c^2 t^2)&s=1\\
(1-\frac{\cos[(s-1)\arctan(\omega_c t)]}{[1+\omega_c^2 t^2]^{(s-1)/2}}\overline{\Gamma}[s-1] & s\neq 1
\end{cases}
\end{align}
where $\overline{\Gamma}[x]=\int_0^{\infty} t^{x-1}e^{-t} dt$. The quantity $\omega_c$ is usually referred to as {\em cutoff frequency} of the bath, whereas $s$ is the so-called 
Ohmicity parameter. The Ohmicity governs the behaviour of the spectrum at low frequency, and identifies three regimes: for $s<1$ the spectrum is referred to as {\em subOhmic}, 
$s=1$ singles out the Ohmic case, and $s>1$ corresponds to a {\em superOhmic} environment. 

As mentioned earlier, our aim is to discriminate between different values of 
the parameters $\alpha$ and $s$ in the classical and quantum case, respectively. 
\section{Learning Task and Models} 
\subsection{Data description}
For a generic state $\ket{\phi}= \beta_1 |0\rangle +\beta_2 |1\rangle$, randomly 
sampled from the Haar distribution in $\mathcal{C}^2$, the associated density 
matrix is $\rho=\ket{\phi}\bra{\phi}$, and the state evolved in a dephasing channels is given by Eq. (\ref{channel}), i.e. 
\begin{align}
     \rho(t,\nu_l) = &
    \begin{pmatrix}
    |\beta_1|^2 & \beta_1\beta_2^* \,\Lambda_l(t, \nu_{l})\\
    \beta_1^*\beta_2\, \Lambda_l(t,{\nu_{l},})& |\beta_2|^2
    \end{pmatrix}
    \label{rho}
\end{align}
where $\Lambda_l(t,\nu^l)$ is the dephasing function  with $l=c,q$ {and we identify the noise parameters $\nu_c=\alpha$ and $\nu_q=s$.}

In order to feed the network, we need to know the density matrix $\rho(t)$ at some time. To this aim we should  consider a reliable tomographic measurement, e.g. a \textit{symmetric informationally complete} probability operator-valued measure (SIC-POVM) i.e. the set $\{\mathcal{M}_k = \ket{M_k}\bra{M_k}, \sum_{k = 1}^{d^2} \mathcal{M}_k= \mathbb{I}, \text{rank} = 1 \}_{k=1}^{d^2} $ where the $\ket{M_k}$ are   $d$-dimensional states with Hilbert-Schmidt product $\Tr(\mathcal{M}_k\mathcal{M}_j)= \tfrac{1 + d\delta_{kj}}{d+1}$ \cite{sicnotes,sicrobustness}.  SIC operators are maximally efficient at estimating the quantum state. For this very reason they are largely employed in tomography and cryptography \cite{sicnotes,sicrobustness} and this makes them good candidates for us as well.

To build our datasets, we leverage the injective mapping between the convex set of density operators and the set of probability distributions over $d^2$  outcomes $p_k(t)=\Tr[\mathcal{M}_k \rho(t)]$ \cite{QBayes}. A quantum state  of a $d$-dimensional system can be thus 
expressed as a $d^2$-dimensional real vector  on a regular simplex $S$ immersed in  $\mathcal{R}^{d^2-1}$,  $\ket{\ket{\textbf{p}_{\Lambda_l(t,\nu_l)}}} = \big(p_1, p_2 \dots p_{d^2}  \big)^T_{\Lambda_l}$, with $l = (c,q)$ (i.e. classical or quantum).
Indeed, the subset of the points representing density matrices in the ${d^2-1}$-simplex  
corresponds to the ${d^2-1}$ generalized Bloch sphere vectors \cite{simplex}. In particular,  we  consider the qubit systems, for which SIC basis can be directly expressed in term of Pauli ones \cite{QBayes}.

For a given number $N$ of initial density matrices, and a fixed {dimensionless}-time windows $D $ of arbitrary width, we consider 
the set of pairs $\{(t_1,t_2), t_1,t_2 \in D \}$. The input states $\textbf{x}_{\nu_l}$ for the networks are devised as follow
\begin{align*}
\big\{t_1, t_2 \big\} & \to
\{\textbf{x}_{\nu_l}\} = \Big\{\pin\bigoplus\pfin \Big\}.
\end{align*}
We create a selection of $m$  arbitrarily chosen   classes of bath parameters $\vec{\nu}_c = (\alpha_1 ...\alpha_m)$ and $\vec{\nu}_q =
(s_1 \dots s_m)$, for the classical and quantum environment, respectively. 
To build our dataset $ \mathcal{D}_{l}$ we collect
and shuffle the inputs $\textbf{x}_{\nu_l}$
for the selected  values of the parameter $\nu_{l}\in \vec{\nu_l}$:
\begin{align}
   \mathcal{D}_{l} = \bigcup_{k=1}^N\,\,\,
   \bigcup_{\nu_l\in\vec{\nu}_l}\,\{ \textbf{x}_{\nu_l},\nu_{l} \}_{\rho_k}\quad l = c,q
\end{align}
where $\rho_k$ is the $k$-th randomly generated density matrix of Eq. \eqref{rho}, and we use the shorthand notation $\nu_{l}\in \vec{\nu_l}$ to indicate the different elements  of the set $\vec{\nu}_l$.
\subsection{Learning Models}
Our task is to learn the function 
\begin{align*}
   f : \, S\bigoplus S &\to \mathcal{R}^m\\
\mathbf{x}_{\nu_l}&\to \textbf{y}_{out}
\end{align*}
mapping the input $\textbf{x}_{\nu_l}$ to the vector $\textbf{y}_{out}$, i.e. the 
probability distribution of the parameter $\nu_l$  over the $m$ classes. 
The class with highest probability is then the estimated value of the bath parameter.
The explicit equations for a feed-forward neural network of $n-2$ hidden layer are given by:
\begin{align}
    h_1 & = \sigma(W_1\mathbf{x}_{\nu_l} + b_1\nonumber) \\
    &\dots\\\nonumber
    h_n &= \sigma(W_{n-1}\,h_{n-1} + b_{n-1}) \\
    y_{out} &= \hbox{softmax}(h_n)\nonumber
\end{align}
where $W_i$ ($i=1,\dots,n-1$) are the weights that the network is going to learn during the training phase, $h_i   $ is the $i$-th network layer,  $b_i$ the bias vectors and $\sigma$ is the activation function, which is the ReLU function $\hbox{ReLU}(x) = \max(0, x)$.
When the network  maps a function from $h_{i-1}\to {h}_i$ with $\text{dim}(h_{i-1})< \text{dim}(h_i)$,  the network is compressing information by mapping the input into a lower space. Otherwise the network is mapping the input to a space of higher dimension.
\begin{figure}[t]
    \centering
    \includegraphics[width =0.91\columnwidth]{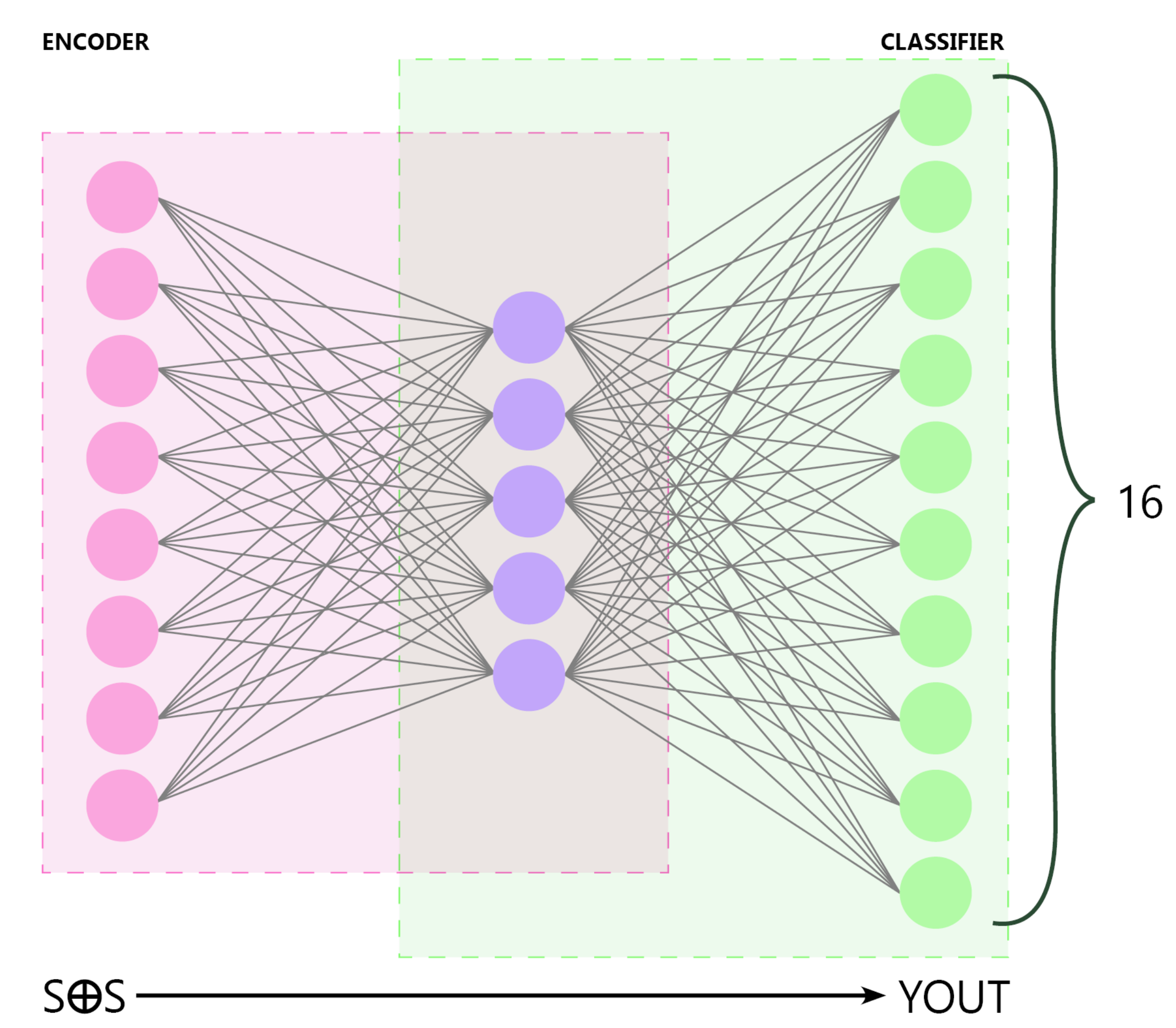}
    \caption{Schematic diagram of our NN-based classification methods.
    We address the use of a linear perceptron (LP), a single inner-layer feed-forward 
    neural network 8-5-16 (in the picture), and a 8-30-30-30-30-30-16 feed-forward NN.}
    \label{f:f1_scheme}
\end{figure}
\par
We use three different models for our analysis: a simple linear perceptron  (LP), which is just an input and output layer, a single inner-layer feed-forward neural network 8-5-16 which is a an encoder with a stacked classifier, and a 8-30-30-30-30-30-16 feed-forward NN.
By construction, the LP model will just be able to provide a linear discrimination  between the classes \cite{Perceptron}.
As usual in machine learning classification tasks, we employ the softmax activation 
function on the output layer
\begin{align}
    \hbox{softmax}(y^{out}_i) = \frac{e^{y_i^{out}}}{\sum_j^{d^2} e^{y_j^{out}}}
\end{align}
which takes the last layer's outputs and turns into a probability distributions 
over the target classes (see Fig. \ref{f:f1_scheme}).{This set-up allows to solve our multinomial logistic regression problem}. As a loss function to train our network, we use the categorical cross-entropy, which quantifies how far is the prediction of the models from the correct class.

In order to assess the predictions made by our classification models, we use two 
performance metrics: accuracy and the macro F1-score. Both are built from the 
confusion matrix $C$, which is a $m\times m$ table where the entry $C_{jk}$ contains the number of events where the $j$-th actual class led to a predicted $k$-th class. 
Given a dataset of  $N$ elements, the first measure is the 
\textit{accuracy}, naturally defined as $A = \tfrac{1}{N}\sum_j C_{jj}$. However, we 
cannot rely on this measure alone, especially when it comes to comparing different models. 
We need a measure that takes into account the imbalances in the class distribution~\cite{FEDECIT}.
In turn, the \textit{precision} is defined as the number of corrected predicted instances 
out of all predicted instances. For each actual class $j$, precision is defined as $P_j = \tfrac{C_{jj}}{\sum_{k}C_{jk}}$. The \textit{recall} is instead defined as the number of correctly predicted instances out 
of the number of true ones. For each predicted class $k$, we have $R_k = \tfrac{C_{kk}}{\sum_{j} C_{jk}}$. Finally, we define the F1 score for each class as the harmonic mean $F1_j = 2\, \tfrac{P_jR_j}{P_j+ R_j}$. The overall metric we use to assess performance is then the \textit{macro averaged F1 score}:
\begin{equation}
    F1= \frac1m \sum_{j=1}^m F1_j,
\end{equation}
which we just refer to as the F1-score hereafter.
\section{Results}
\subsection{Datasets preparation}
For each dataset, we employ  $N=2500$ Haar-generated density matrices  
undergoing the dephasing channel of Eq. (\ref{rho}). Then, we create 1000 time-ordered 
couples $\{(t_1,t_2)\}$ using  110 random times, uniformly distributed in the selected 
time window D. In addition, we set the following physical parameters: 
the frequency parameters for colored noise lies in the interval $[\gamma_1,\gamma_2]=[10^{-4}, 10^4 ]$, whereas the cutoff frequency for the spin-boson model is set to $w_{c}$ = 1. All the functions are implemented using a Python library for real and complex floating-point 
arithmetic with arbitrary precision \cite{mpmath}. 

We create different datasets for our experiments as follows. At first, we want to test 
how our models work with noiseless data. This is relevant in order to gauge
the absolute performance of our learning models. Then, we introduce Gaussian measurement 
noise into our data $\mathbf{x}_{\nu_l}$. In particular, we assume that each $p_i(t)$ 
is affected by a random Gaussian perturbation, taken from the normal distribution  
$\mathcal{N}$(0,\,0.01).

We then focus on two different classification tasks. First, for noiseless and noisy data, we establish $m=16$ classes $\vec{\nu_l}$. Then, we reshape our noisy data into 2 coarse-grained classes  $\vec{\mu}_l = (\nu_l \leq 1,\nu_l >1  )$ with $l=c,q$. The two-class datasets allow 
us to assess whether the network is able to classify data according to a leading physical 
property of the environment, i.e. being more or less 'colored' for classical noise or
being sub/super-Ohmic for the quantum bath.
\begin{figure}[t]
\begin{center}
    \includegraphics[width =0.75\columnwidth]{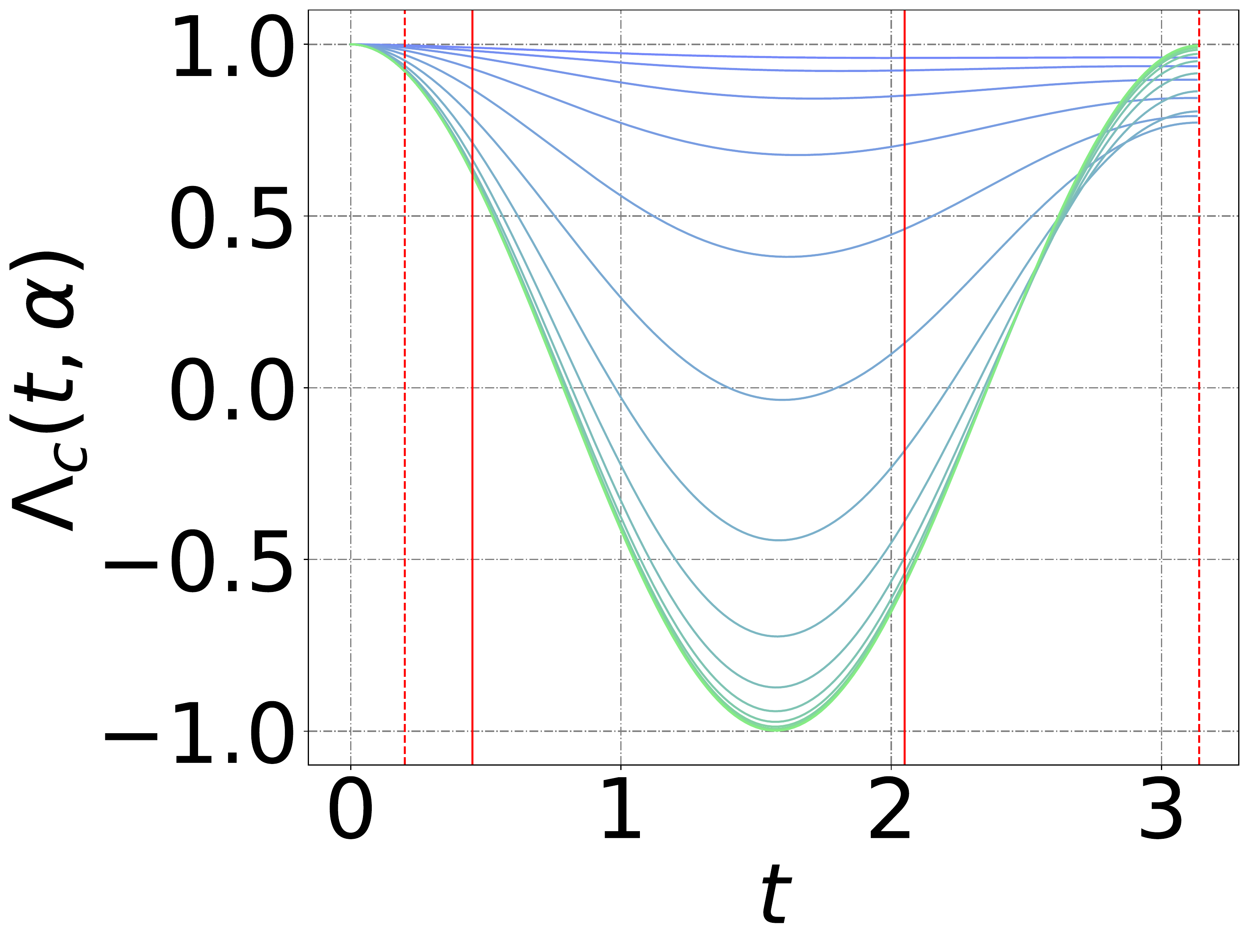}
    \includegraphics[width =0.75\columnwidth]{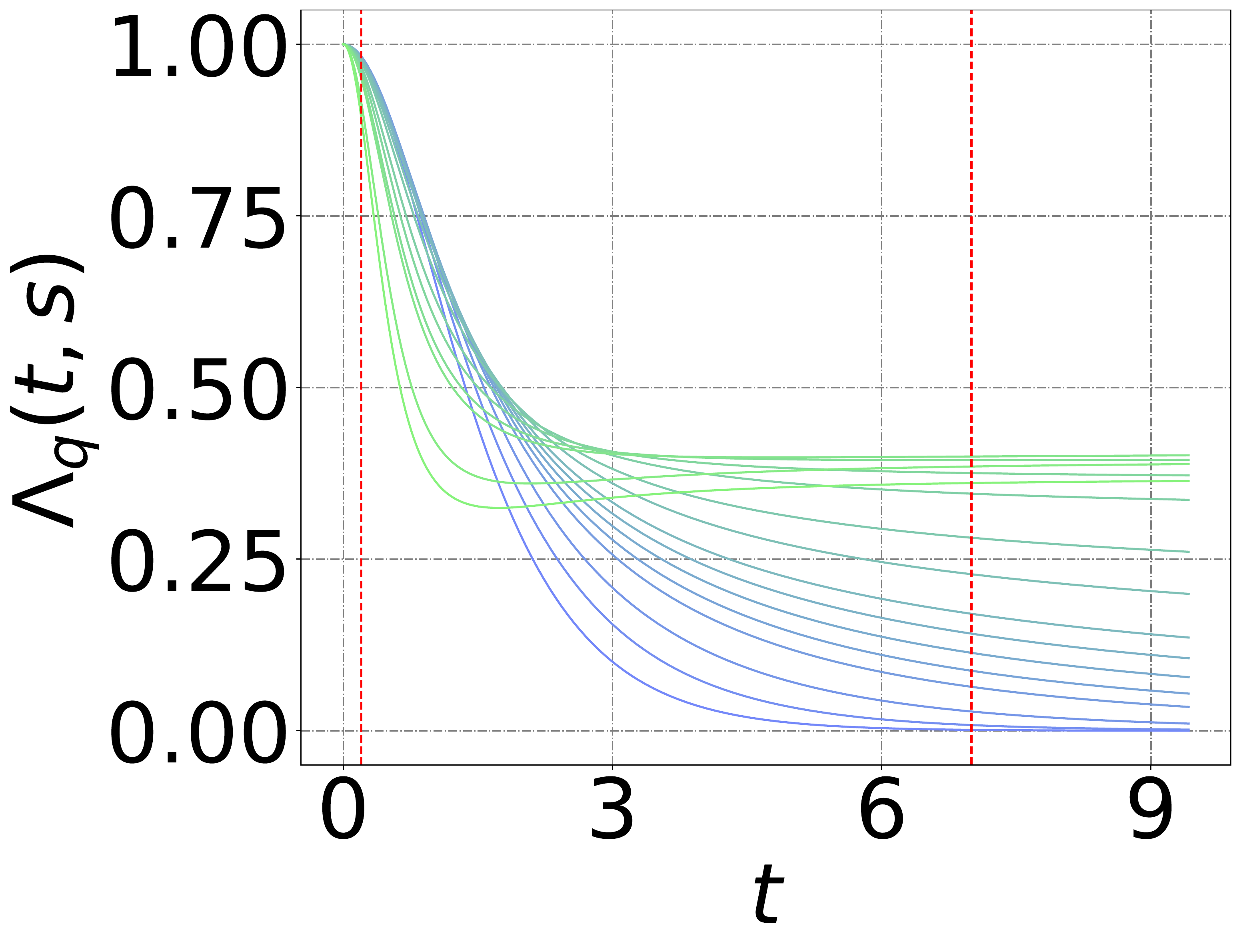}
    \end{center}
  \caption{The dephasing terms $\Lambda_c(t,\alpha)$ and $\Lambda_q(t,s)$ for classical 
  colored noise (upper panel) and the Ohmic quantum noise (lower panel). For classical noise we show $\Lambda_c(t,\alpha)$ for $\alpha$ in the range $[0.5,2]$ (from top to bottom in the left part of the figure). For quantum noise we show $\Lambda_q(t,s)$ for $s$ in the range $[0.1,3]$ (from bottom to top in the right part of the figure). The vertical lines in both panels refer to the time windows $D$ used in the data generation stage (chosen to contain the relevant dynamics of the system). For classical noise we have $D=[0.2,3.14]$ (dashed vertical lines) 
  in the noiseless case and $D=[0.4,2.15]$ (solid) for the noisy 
scenario. For quantum noise we have $D=[0.2,7]$ in both cases. }
\label{f:f2data}
\end{figure}  
\par
All datasets consist of $15.3 \cdot 10^5$ shuffled-input training data, and $3.6\cdot 10^5$ 
shuffled-input for (both) validation and test. The time windows are chosen to cover the 
relevant dynamics of the system, they are $D=[0.2, 3.14]$ for noiseless colored data, 
$D=[0.4, 2.15]$ for noisy colored data, and $D=[0.2, 7]$ for the bosonic bath (noiseless 
and noisy case, see Fig. \ref{f:f2data}).
We use pure initial states, again to test the model in ideal conditions, and mixed ones with purity $\Tr[\rho(0,\nu_l)^2]=0.72$, to mimic realistic noisy conditions. The mixed states are obtained 
by depolarizing the initial pure states.
\subsection{Noiseless measurement data}
The noise-free scenario aims to test whether, and to which extent, a LP or a NN are 
able to solve the classification problem using 16 classes for the dephasing parameters. 
For this task the NN architecture is set to 8-5-16 with RElu activation. For both NN and LP we use the following specifications: batchsize = 300 for training, the dataset is 
split in test, train, and validation, with the validation dataset monitoring the 
early-stopping  function, set at one, to avoid overfitting \cite{earlyS}.

\begin{figure}[t]
    \centering
    \includegraphics[width = 0.95\columnwidth]{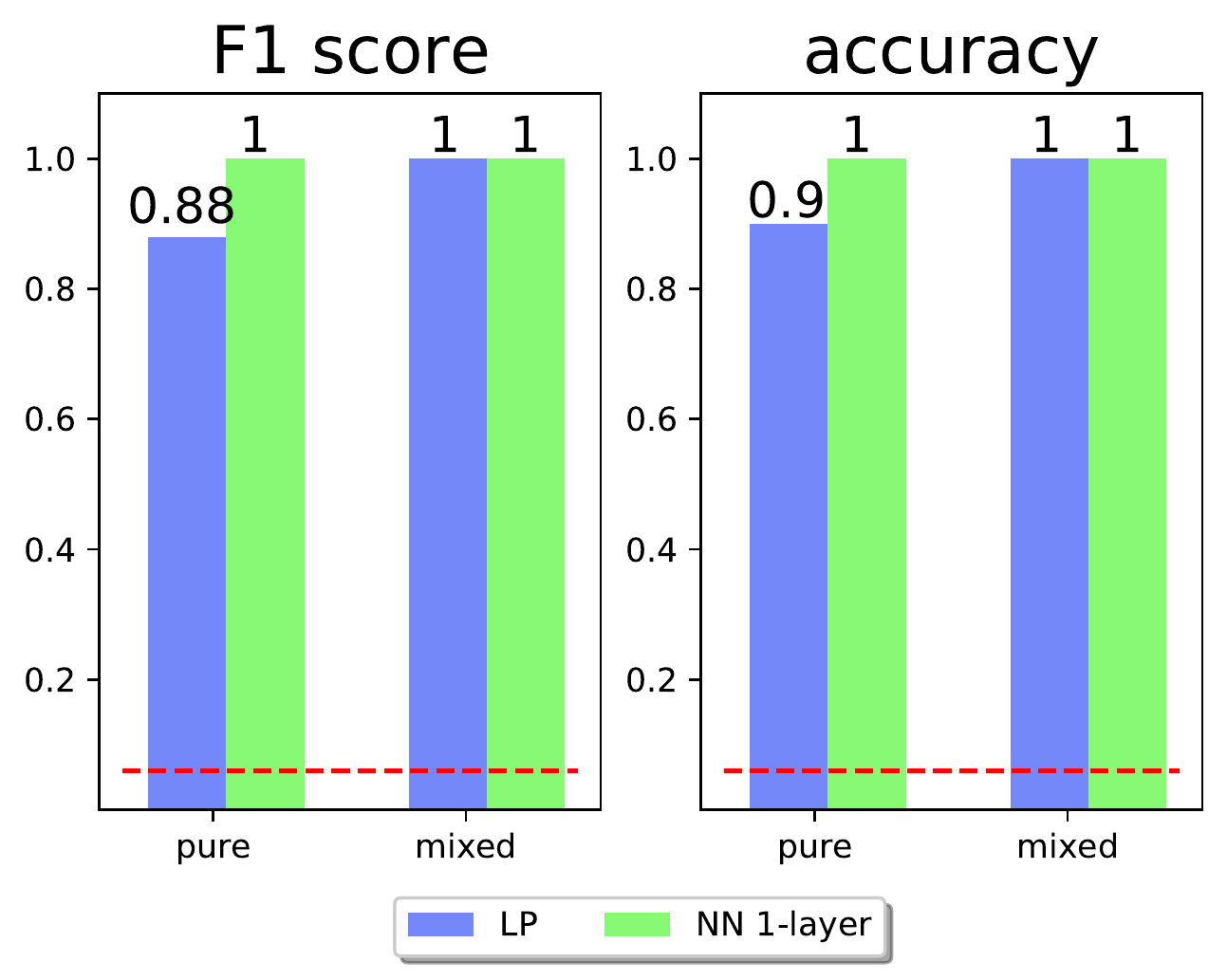}
    \caption{
    F1-score and accuracy for Ohmic dephasing classification for the 16-classes problem. For the linear projector, the F1-score and accuracy values are [0.88,0.9] and [1,1] for pure and mixed probes, respectively. The NN model achieves perfect classification for 
    both type of probes.
 The red dashed line marks the 'random guess' for F1-score and accuracy.     }
    \label{boson}
\end{figure}

For classical colored noise the problem is linear.  Our experiments confirm that the 
problem may be easily solved with both (LP and NN) models, achieving $100\, \%$ accuracy 
(and F1-score) with either pure or mixed initial states.  In the Ohmic case, we observe
 a difference in the performance of the two classification models. The LP model cannot achieve perfect classification using pure initial state states, while a single inner layer NN 
achieves perfect environment discrimination. 
Results are summarized in Fig. (\ref{boson}). We anticipate that the small  differences 
between the performance of the different models will be amplified in the realistic scenario, 
i.e., when measurement noise is added.  

Additionally, we want to understand how our network generalizes with respect to novel 
data, by training and testing the network with data coming from different regions of the Bloch 
sphere, e.g. states with a constraint in the component along the $z$-axis.
The train and the test datasets are given by
\begin{align*}
  & D_{\nu_l} = \{\textbf{x}_{\nu_l}:|\Tr \left[\sigma_z\,\rho(0,\nu_l)\right]|< 0.7  \} \quad \hbox{train \& validation}\\
  & \overline{D}_{\nu_l} = \{\textbf{x}_{\nu_l} : 0.7 <|\Tr\left[\sigma_z\rho(0, \nu_l)\right]|< 0.8 \}\quad \hbox{test}
\end{align*}
respectively. In this way we can ensure that the test states are always far enough 
and separated from our train/validation set (in the Euclidean sense).  In this case, we 
use  the 8-5-16 architecture and the experiment aims to assess whether the network is truly 
inferring the bath  parameters from the $\{\textbf{x}_{\nu_l}\}$ inputs, or we are   
just over-fitting.  Results for the single-layer NN confirm that our model is able 
to classify novel  data. Indeed, we obtain perfect scoring for accuracy and macro-F1 (both 100\%).
This provides further evidence in favor of our basic hypothesis: 
the network is learning a function able to catch the hallmark of the baths. 

Finally, we also consider a situation in which we train the network over initial pure states and test it with 
mixed initial states. Specifically, the datasets are:
\begin{align*}
D_{\nu_l}= \{\textbf{x}_{\nu_l} \: \Tr\left[\rho^2(0,\nu_l)\right] & = 1\} \quad \hbox{train}\\
\overline{D}_{\nu_l}= \{ \textbf{x}_{\nu_l} \: \Tr(\rho^2(0,\nu_l)) & = 0.72\} \quad \hbox{test}
\end{align*}
The test dataset is obtained by depolarizing the initial pure states. We consider the 
16 classes problem, and use the 8-5-16 model. Batch-size is equal to 300 and early stopping is 
set to one. The above single-layer NN is indeed able to discriminate the bath parameters, 
obtaining an accuracy $\sim 100\%$  for the color $\alpha$ and $\sim 88\%$ 
for the Ohmicity $s$.

\subsection{Noisy measurement data}
If data are affected by Gaussian noise, the performance of LP and NN models degrades. In particular, LPs fail to classify parameters due to the strong nonlinearity of the problem.
For the 16-classes problem, we employ both the 8-5-16 NN model and 8-30-30-30-30-30-16 NN architecture. Batchsize and early stopping parameters are left unchanged. Similar 
architectures have been exploited in other applications facing noisy datasets \cite{NMR,adrifede}. 

\begin{figure}[t]
    \centering
    \includegraphics[width = 0.95\linewidth]{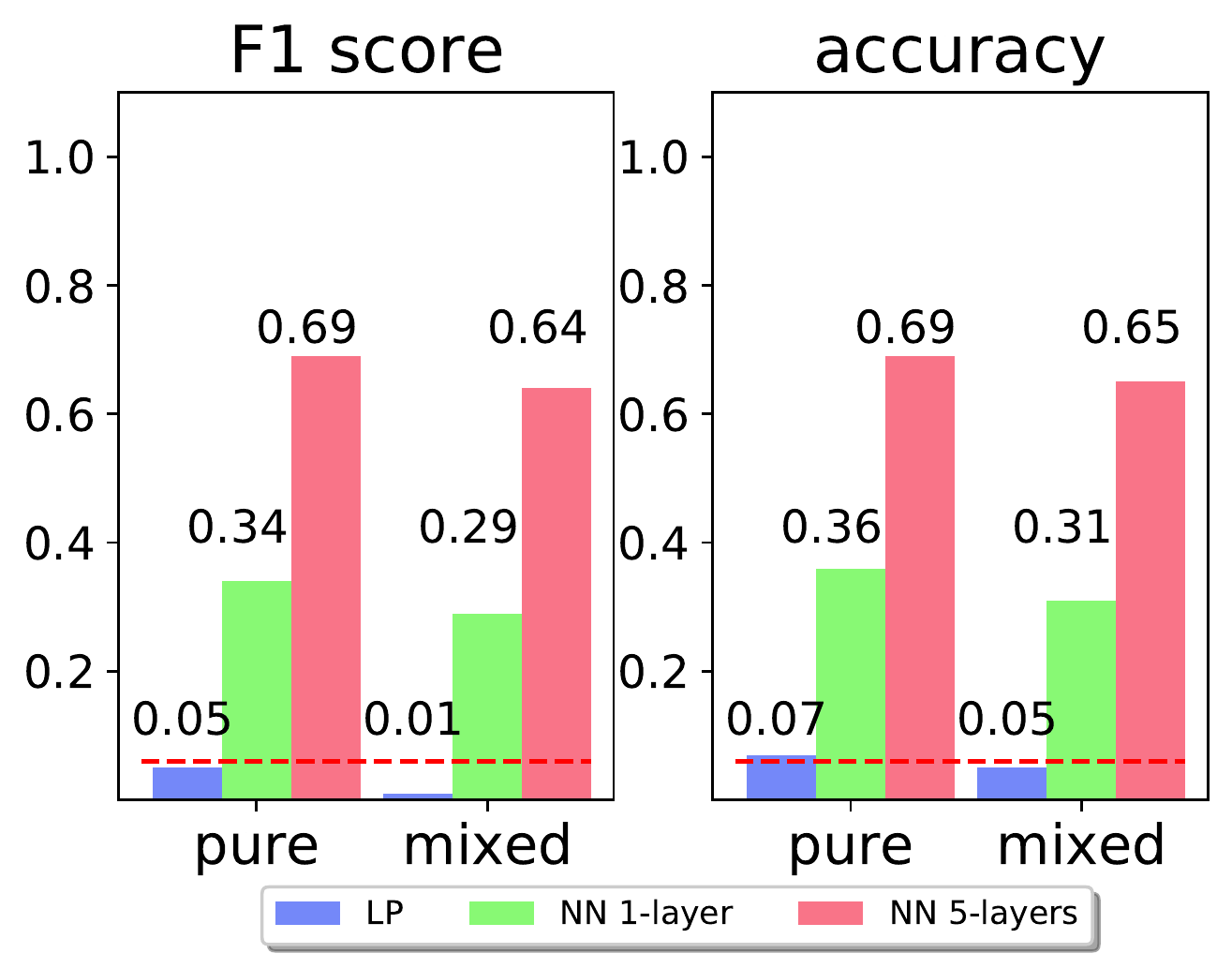}
    \caption{F1-score and accuracy for noisy classification for colored dephasing, 16 classes. For this task the 3 different models increase in complexity stepwise, offering us a global view of the increasing level of difficulty of the task.
    The LP model  F1-score and accuracy values are [0.05, 0.07] for pure initial states and 
    [0.01, 0.05] for mixed ones. The single  layer NN model values  are [0.34, 0.36] for initial pure states and [0.29, 0.31] for mixed. With the 5-layers model NN we achieve  [0.690, 0.695] for pure states
    and [0.64, 0.65] for mixed ones. The red dashed line marks the 'random guess' for F1-score and accuracy.}
    \label{colorednoise16}
\end{figure}

Results for the colored noise classification case are shown in Figure \ref{colorednoise16}. 
LP model apparently fails at classifying noise, reaching at most  the order of the random 
guess ($\sim$ 0.06). The non-linear NN architectures can instead achieve higher values of 
F1-score and accuracy by increasing the number of layers. We found that for 16 classes problem, 
the noisy case requires wider inner layers. In other words, data are mapped into a space 
of higher dimension than their initial one. In order to illustrate the model statistics, we  
report the confusion matrix for the 16-classes colored classical noise (mixed initial state) 
in Appendix \ref{app1}. The classification task is instead challenging in the Ohmic noisy case, where the best performer is the 5 inner-layers model and is able to achieve an accuracy of
0.25, corresponding to a F1-score equal to 0.22.

\subsubsection{A two-class classification problem}
In order to investigate the problem from a physical driven perspective, we also consider the case
where the network is given the problem to distinguish between the
$\alpha\lessgtr 1$ regimes for the classical noise and
sub/super-Ohmic dephasing ($s\lessgtr 1$) for the quantum bath. 

\begin{figure}[t]
    \includegraphics[width = 0.9 \columnwidth]{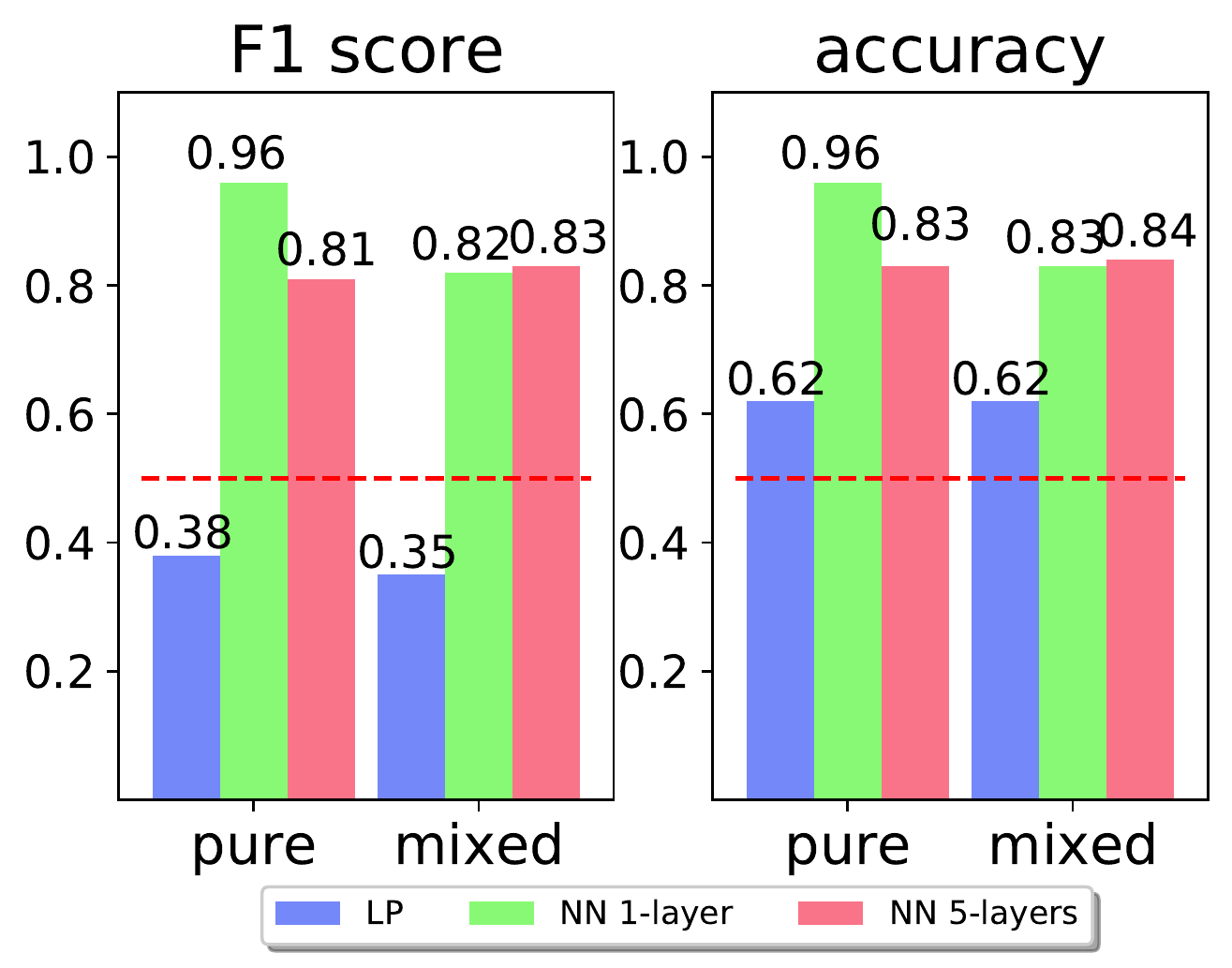} 
    \includegraphics[width = 0.9 \columnwidth]{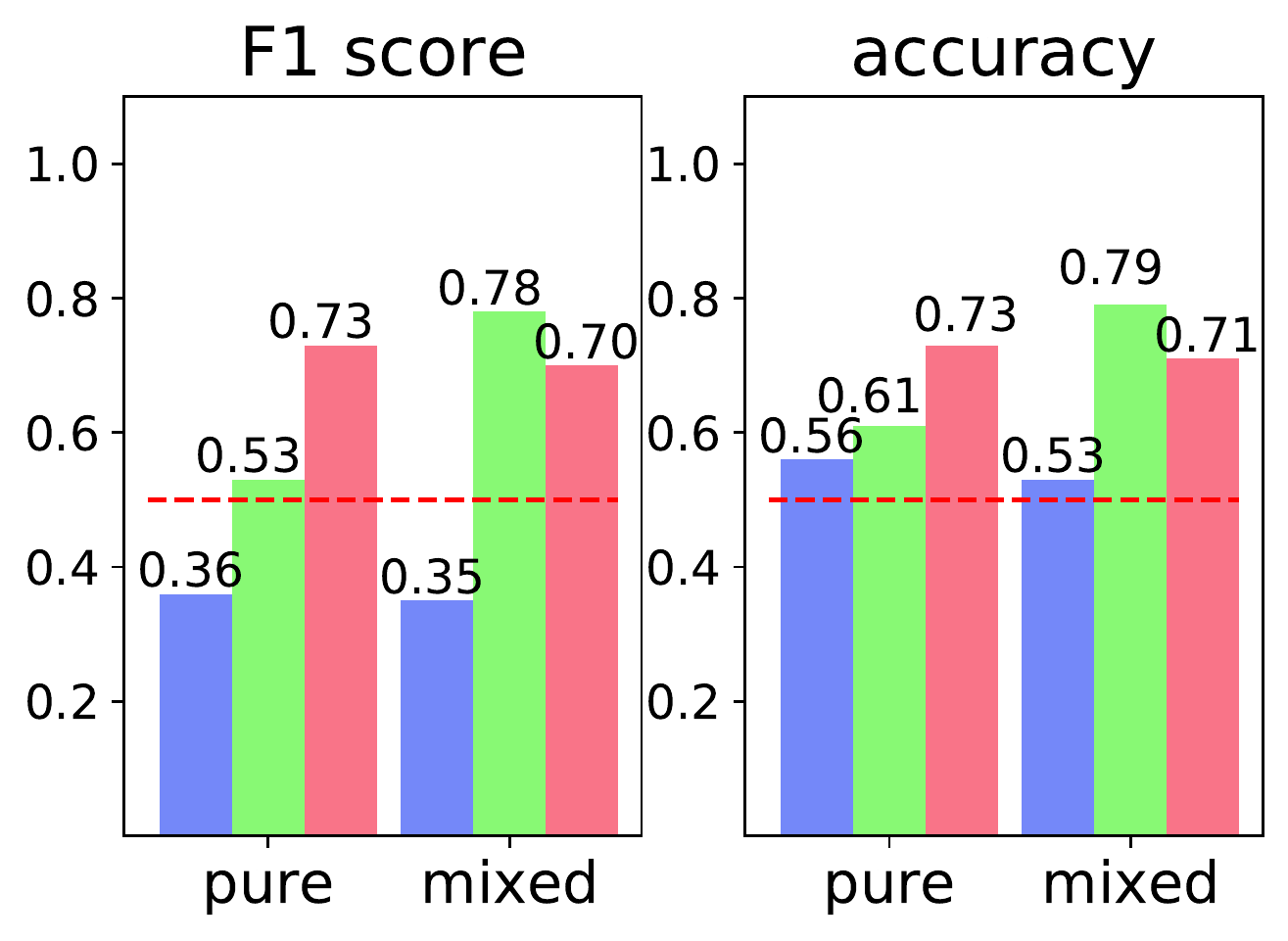} 
    \caption{F1-score and accuracy for the two-class classification problem involving 
    colored (upper panels) and Ohmic noise (lower panels).  For colored noise, 
    the largest values of F1-score and accuracy are obtained by the single-layer NN (0.96\%) 
    for pure states, while for mixed states the two models are almost equivalent.
    For Ohmic dephasing, we obtain the best results  with the 5-layers NN with pure states 
    and with the single-layer NN for mixed-states. The red  dashed  line  denotes  
    the  'random  guess' threshold  for  F1-score  and accuracy.}
    \label{twocl}
\end{figure}

Results are illustrated in Fig. \ref{twocl}, which shows how the problem is non-linear and highly dependent on the classical or quantum nature of the environment. For classical colored noise
with pure initial states, the  best performer is the single-layer NN, able to achieve  $96\% $ in  both accuracy and F1. For mixed states, the corresponding values of F1 and accuracy are  $83\%$ and $82\%$, just a $1\%$ less with respect to the 5-layers NN, however obtained with a simpler architecture. Indeed,  
the  model deals with the noise perturbations efficiently. This is another evidence that the network captures the physical fingerprint of the environment parameters. 

For the quantum bath with mixed initial states, a $8\%$ advantage is gained by the 1-layer model. 
On the contrary, for  pure initial states the 5-layers model leads to a higher accuracy and F1 score. This behavior may be understood by considering the data sparseness. Indeed,
if we add a L1 regularization term \cite{regularization} the network may reach up to $81\%$ accuracy and $80\%$ for F1-score. Overall, the single-layer network turns out as the most efficient model for tackling the two-class task.

\section{Conclusions}
In this paper, we have analyzed the use of neural networks to classify the noise 
parameter of dephasing channels. We have considered channels originating either 
from classical $1/f^{\alpha}$ noise or from the interaction with a bath of quantum 
oscillators. Our strategy involves the use of a single qubit to probe the channel 
and requires, to train the classifying network, the knowledge of the qubit density 
matrix at just two random instants of time within a fixed time-window.

At first, we have shown that the qubit dephasing problem may be re-casted into a 
linear discrimination task,  solvable with minimal resources, e.g. basic-vanilla ML 
models. In particular, we have shown that our network is able to exactly classify 
spectral parameters into 16 classes using noiseless data. Then, in order to assess the 
performance in a more realistic scenario, we added Gaussian noise to data.
For the 16 classes task linearity is lost for both colored and Ohmic dephasing. 
The classical noise classification is still feasible with a 5-layer NN model, leading to 
a relatively high accuracy, whereas quantum dephasing classification is prone to fail.
We also trained the network to discriminate between to macro-classes, involving either $\alpha\lessgtr 1$ or $s\lessgtr 1$.  In these cases the single layer model  outpaces  the 5-layer one, reaching levels of F1 and accuracy of 96\% for the colored and 79\% for the Ohmic environment.

Our results  confirm the feasibility  of the approach, i.e. the use of neural networks 
in classifying the environmental parameters of single-qubit dephasing channels. This is 
remarkable, in view of the small amount of data (the state at two random instants of time) 
needed to use the network. Our results 
show that, in general, bosonic baths are more demanding to discriminate compared to classical 
$1/f^{\alpha}$ noise, and that NN may not be able to achieve multiclass classification of quantum noise. For the simpler task of two-class classification a single-layer NN is able to effectively discriminate noise, either classical or quantum ). 

Our results show that quantum probing may be effectively enhanced by the use of NNs, which 
provide a way to effectively extract information from tomographic data, and pave the way for 
future investigations devoted to more advanced and in-depth architectures for realistic 
use-case, e.g. to unveil the reason under the bosonic bath hardships.
\acknowledgments
MGAP is member of INdAM-GNFM. FB is a member of the Bocconi Institute for Data Science and Analytics (BIDSA) and the Data and Marketing Insights (DMI) unit.
\bibliography{qdbib}
\appendix 
\section{16-classes confusion matrices}\label{app1}
\begin{figure}[h!]
    \includegraphics[width=0.9\columnwidth]{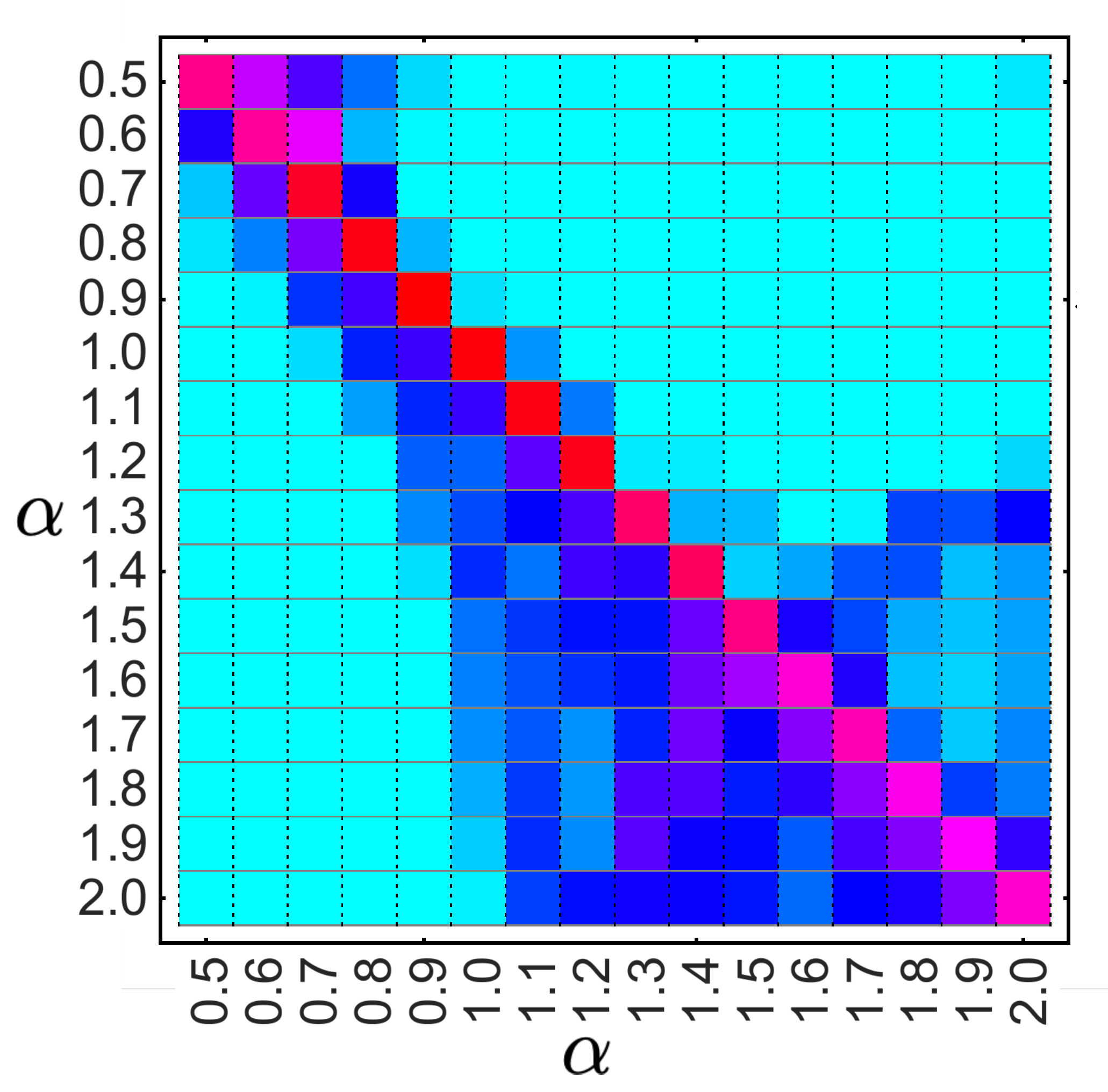}
   \caption{Confusion matrix for a 16-class classification model involving mixed states 
   undergoing a colored noise dephasing and subject to Gaussian noise in the tomographic stage (red denotes higher values and light blue lower ones). 
   The higher contributions to the confusion correspond to the region $\alpha = [1.7, 2]$ where the dephasing factors become closer and  closer, so as to make very difficult for the network to find optimal separating hyperplanes.}
    \label{fig:confusions}
\end{figure}

For the sake of completeness, we report here (see Fig. \ref{fig:confusions}) an example
of confusion matrix for a 16-classes classification task involving classical colored 
noise and initially mixed states in the presence of noise. Similar patterns may be observed 
for pure initial states. The good accuracy level is witnessed by the higher values 
of the diagonal entries, whereas the good F1 score is achieved because the off-diagonal 
elements are smaller. Notice that the accuracy starts to drop from class 
$\alpha = 1.2$ and the larger drop may be seen in the range $[1.7, 2]$, where 
the off-diagonal entries takes a non-negligible value. This brings a geometrical 
consideration inspired by the behavior of the dephasing factor $\Lambda_c(t,\alpha)$: 
when noise is added, the data are no more well separated. The smaller is the difference 
$|\Lambda_c(t,\alpha_j) - \Lambda_{c}(t,\alpha_{j+1})| $, the harder is for the network 
to identify an optimal separating hyperplane between classes $j$ and $j+1$. This is 
especially true for the  classes $\alpha \in [1.7,2]$. 
\end{document}